\documentclass[twocolumn,prb,showpacs]{revtex4}

\usepackage{graphicx}
\usepackage{dcolumn}
\usepackage{bm}
\usepackage{amsmath}

\begin{document}

\title{Anomalous physical properties of underdoped weak-ferromagnetic superconductor RuSr$_2$EuCu$_{2}$O$_{8}$}

\author{B. C. Chang}
\author{C. Y. Yang}
\author{H. C. Ku}%
\email{hcku@phys.nthu.edu.tw} \affiliation{Department of Physics,
National Tsing Hua University, Hsinchu 300, Taiwan, Republic of
China}

\author{J. C. Ho}
\affiliation{Department of Physics, Wichita State University,
Wichita, Kansas 67260-0032, U.S.A.}

\author{C. B. Tsai}
\author{Y. Y. Chen}
\affiliation{Institute of Physics, Academia Sinica, Taipei 115,
Taiwan, Republic of China}

\author{D. C. Ling}
\affiliation{Department of Physics, Tamkang University, Tamsui 251,
Taiwan, Republic of China}

\date{\today}


\begin{abstract}
Similar to the optimal-doped, weak-ferromagnetic (WFM induced by
canted antiferromagnetism, T$_{Curie}$ = 131 K) and superconducting
(T$_{c}$ = 56 K) RuSr$_{2}$GdCu$_{2}$O$_{8}$, the underdoped
RuSr$_{2}$EuCu$_{2}$O$_{8}$ (T$_{Curie}$ = 133 K, T$_{c}$ = 36 K)
also exhibited a spontaneous vortex state (SVS) between 16 K and 36
K. The low field ($\pm$20 G) superconducting hysteresis loop
indicates a weak and narrow Meissner state region of average lower
critical field B$_{c1}^{ave}$(T) = B$_{c1}^{ave}$(0)[1 -
(T/T$_{SVS}$)$^{2}$], with B$_{c1}^{ave}$(0) = 7 G and T$_{SVS}$ =
16 K. The vortex melting transition (T$_{melting}$ = 21 K) below
T$_{c}$ obtained from the broad resistivity drop and the onset of
diamagnetic signal indicates a vortex liquid region due to the
coexistence and interplay between superconductivity and WFM order.
No visible jump in specific heat was observed near T$_{c}$ for Eu-
and Gd-compound. This is not surprising, since the electronic
specific heat is easily overshadowed by the large phonon and
weak-ferromagnetic contributions. Furthermore, a broad resistivity
transition due to low vortex melting temperature would also lead to
a correspondingly reduced height of any specific heat jump. Finally,
with the baseline from the nonmagnetic Eu-compound, specific heat
data analysis confirms the magnetic entropy associated with
antiferromagnetic ordering of Gd$^{3+}$ (J = S = 7/2) at 2.5 K to be
close to $\it{N_{A}k}$ ln8 as expected.

\end{abstract}

\maketitle

\section{Introduction}

Anomalous physical properties have been observed recently in the
weak-ferromagnetic (WFM induced by canted antiferromagnetism) and
high-T$_{c}$ superconducting RuSr$_{2}$RCu$_{2}$O$_{8}$ system
(Ru-1212 with R = Sm, Eu, Gd, and Y) having a tetragonal
TlBa$_{2}$CaCu$_{2}$O$_{7}$-type
structure.\cite{p1,p2,p3,p4,p5,p6,p7,p8,p9,p10,p11,p12,p13,p14,p15,
p16,p17,p18,p19,p20,p21,p22,p23,p24,p25,p26,p27,p28,p29,p30,p31,p32,
p33,p34,p35,p36,p37,p38,p39,p40,p41,p42,p43,p44,p45,p46,p47,p48}
Possible superconductivity was also reported in Ca-substituted WFM
compounds RuCa$_{2}$RCu$_{2}$O$_{8}$ (R = Pr-Gd).\cite{p49,p50,p51}
The weak-ferromagnetism in these strongly-correlated electron
systems originates from the long range order of Ru moments in the
RuO$_{6}$ octahedra due to a strong
Ru-4$\it{d}_{xy,yz,zx}$-O-2$\it{p}_{x,y,z}$ hybridization with a
Curie temperature T$_{Curie} \sim$ 131 K. A G-type antiferromagnetic
order probably occurs with Ru$^{5+}$ moment $\mu$ canted along the
tetragonal basal plane, even through the small net spontaneous
magnetic moment $\mu_{s}$ $\ll \mu$(Ru$^{5+}$) is too small to be
detected in neutron diffraction.\cite{p4,p5,p9,p10,p22} The Ru
valence of 4+ and 5+ was determined from x-ray absorption near edge
measurements. \cite{p23,p52}

With its quasi-two-dimensional CuO$_{2}$ bi-layers separated by a
rare earth layer in the Ru-1212 structure,
RuSr$_{2}$GdCu$_{2}$O$_{8}$ has the highest resistivity-onset
temperature T$_{c} \sim$ 60 K among different Ru-1212 compounds.
\cite{p1,p2,p4,p5,p31} A broad resistivity transition width
$\Delta$T$_{c}$ = T$_{c}$(onset) - T$_{c}$(zero) = T$_{c}$ -
T$_{melting}$ $\sim$ 15-20 K is most likely a consequence of
coexistence and interplay between superconductivity and WFM order.
The diamagnetic signal is observed only near T$_{melting}$ instead
of T$_{c}$, and a reasonable large Meissner signal can be detected
only in zero-field-cooled (ZFC) mode.\cite{p47} Lower T$_{c} \sim$
40 K and 12 K were observed for Eu-compound and Sm-compound,
respectively.\cite{p12,p18} No superconductivity can be detected in
RuSr$_{2}$RCu$_{2}$O$_{8}$ (R = Pr, Nd),\cite{p3,p16} while a
superconducting RuSr$_{2}$YCu$_{2}$O$_{8}$ phase is stable only
under the high pressure.\cite{p21,p26}

Interest of the current work stimulates from a recent report of
spontaneous vortex state (SVS) between 30 K and 56 K in
RuSr$_{2}$GdCu$_{2}$O$_{8}$.\cite{p47} However, the compound
undergoes a low temperature antiferromagnetic ordering arising from
Gd$^{3+}$ at 2.5 K. To avoid this complication, isostructural
RuSr$_{2}$EuCu$_{2}$O$_{8}$ with nonmagnetic-Eu$^{3+}$ ions was
chosen as a prototype material in this study to evaluate the
anomalous magnetic, transport, calorimetric properties and
$\it{d}$-wave nature near and below T$_{c}$ = 36 K. The calorimetric
data were further used as a basis in elucidating the magnetic
entropy associated with the Gd$^{3+}$ ordering.

\section{experimental}
Stoichiometric RuSr$_{2}$RCu$_{2}$O$_{8}$ samples were synthesized
by solid-state reactions. High-purity RuO$_{2}$ (99.99 $\%$),
SrCO$_{3}$ (99.9 $\%$), R$_{2}$O$_{3}$ (99.99 $\%$) (R = Pr, Nd, Sm,
Eu, and Gd), and CuO (99.9 $\%$), in the nominal composition ratios
of Ru:Sr:R:Cu = 1: 2: 1: 2, were well mixed and calcined at
960$^{\circ}$C in air for 16 hours. The calcined powders were then
pressed into pellets and sintered in flowing N$_{2}$ gas at
1015$^{\circ}$C for 10 hours to form RuSr$_{2}$RO$_{6}$ and
Cu$_{2}$O precursors. This step is crucial in order to avoid the
formation of impurity phases. The N$_{2}$-sintered pellets were
heated at 1060$^{\circ}$C in flowing O$_{2}$ gas for 10 hours to
form the Ru-1212 phase, then oxygen-annealed at a slightly higher
1065$^{\circ}$C for 7 days and slowly furnace-cooled to room
temperature with a rate of 15$^{\circ}$C per hour.\cite{p47}

Powder x-ray diffraction data were collected with a Rigaku Rotaflex
18-kW rotating-anode diffractometer using Cu-K$_{\alpha}$ radiation.
Four-probe electrical resistivity measurements were performed with a
Linear Research LR-700 ac (16Hz) resistance bridge from 2 K to 300
K. Magnetic susceptibility and magnetic hysteresis measurements from
2 K to 300 K in low applied magnetic fields were carried out with a
Quantum Design $\mu$-metal shielded MPMS2 superconducting quantum
interference device (SQUID) magnetometer. Calorimetric measurements
were made from 1 K to 70 K by using a thermal-relaxation
microcalorimeter. A mg-size sample was attached with a minute amount
of grease to a sapphire holder to ensure good thermal coupling. The
sample holder had a Cernox temperature sensor and a Ni-Cr alloy film
heater. The holder was linked thermally to a copper block by four
Au-Cu alloy wires. The temperature of the block could be raised in
steps but held constant when a heat pulse was applied. Following
each heat pulse, the sample temperature relaxation rate was
monitored to yield a time constant $\tau$. The total heat capacity
was calculated from the expression c = $\kappa\tau$, where $\kappa$
is the thermal conductance of Au-Cu wires. The heat capacity of the
holder was measured separately for addenda correction. The molar
specific heat of the sample was then obtained from C = (c -
c$_{addenda}$)/(m/M) with m and M being the sample's mass and molar
mass, respectively.

\section{results and discussion}

\begin{figure}
\includegraphics{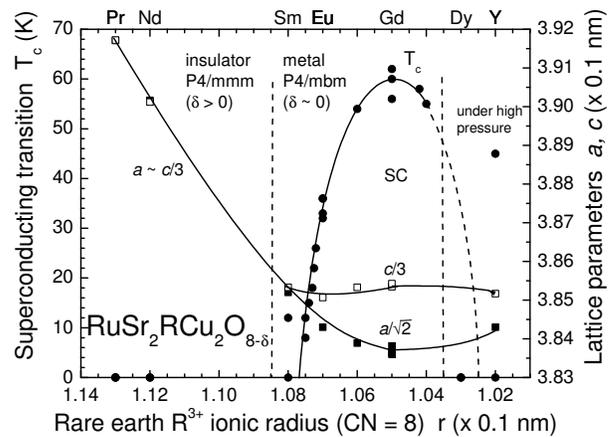}
\caption{\label{fig1} The variation of superconducting transition
T$_{c}$ and tetragonal lattice parameters $\it{a}$, $\it{c}$ with
rare earth ionic radius R$^{3+}$ (coordination number CN = 8) for
RuSr$_{2}$RCu$_{2}$O$_{8-\delta}$ system (R = Pr-Y).}
\end{figure}

Figure 1 summarizes structural and superconducting properties, as a
function of R$^{3+}$ ionic radius r (coordination number CN = 8), of
various RuSr$_{2}$RCu$_{2}$O$_{8-\delta}$ system (R = Pr-Y). T$_{c}$
decreases from a maximum value of 60 K for optimal-doped Gd (r =
0.105 nm) to 36 K for underdoped Eu (r = 0.107 nm), and $<$ 10 K for
Sm (r = 0.108 nm). Larger rare earth ions of Nd (0.112 nm) and Pr
(0.113 nm) lead to a metal-insulator transition. Powder x-ray
Rietveld refinement study indicates that the insulating phase is
stabilized in the undistorted tetragonal phase (space group P4/mmm)
with a larger lattice parameter $\it{a} \sim$ 0.390-392 nm, which
gives a reasonable Ru$^{5+}$-O bond length of d $\sim$ 0.197 nm if
the oxygen content is slightly deficient ($\delta >$ 0). On the
other hand, the metallic phase with smaller rare earth ions can be
stabilized in the full-oxygenated ($\delta \sim$ 0), distorted
tetragonal phase (space group P4/mbm) with smaller
$\it{a}$/$\sqrt{2}$ $\sim$ 0.383-0.385 nm but still a reasonable
Ru-O bond length through RuO$_{6}$ octahedron rotation.

Indeed, the powder x-ray diffraction pattern for the oxygen-annealed
RuSr$_{2}$EuCu$_{2}$O$_{8-\delta}$ sample indicates single phase
with tetragonal lattice parameters of $\it{a}$ = 0.5435(5) nm and
$\it{c}$ = 1.1552(9) nm. A Raman scattering peak of 265 cm$^{-1}$
indicates that the A$_{1g}$ mode symmetry belong to a P4/mbm instead
of P4/mmm group. Accordingly, with RuO$_{6}$ octahedra rotation
angle $\theta \sim$ 14$^{\circ}$ around the c-axis and oxygen
parameter $\delta \sim$ 0,\cite{p10} Rietveld refinement analysis
with a small residual error factor R = 5.31$\%$ yields a reasonable
Ru-O bond lengths d = ($\it{a}$/2$\sqrt{2}$)(1 - sin$^{2}
\theta$)$^{-1/2}$ = 0.198 nm. It is close to the minimum calculated
bond length d(Ru$^{5+}$-O) of 0.197 nm.\cite{p10}

Figure 2 shows the temperature dependence of field-cooled (FC) and
zero-field-cooled (ZFC) volume magnetic susceptibility
4$\pi\chi_{V}$ at 1-G for bulk and powder
RuSr$_{2}$EuCu$_{2}$O$_{8}$ samples. Weak-ferromagmagnetic ordering
occurs at T$_{Curie}$ = 133 K. Similar to
RuSr$_{2}$GdCu$_{2}$O$_{8}$,\cite{p47} this Eu-compound has its
electrical resistivity data, which are also included in Fig. 2,
exhibiting a non-Fermi-liquid-like behavior above T$_{Curie}$. The
linearly temperature-dependant values of 10.0 m$\Omega$ cm at 300 K
and 5.5 m$\Omega$ cm at 160 K give an extrapolated value of 2.6
m$\Omega$ cm at 0 K, yielding a ratio $\rho$(300 K)/$\rho$(0 K) of
3.9. Below T$_{Curie}$, a T$^{2}$ behavior prevails. The onset of
deviation at 36 K from such a temperature dependence is taken as the
superconducting transition temperature T$_{c}$. The melting
temperature of superconducting vortex liquid is assigned to
T$_{melting}$ = 21 K, where resistivity reaches zero.\cite{p47} The
broad transition width of 15 K is the common feature for all
reported Ru-1212 compounds. It indicates that the superconducting
Josephson coupling along the tetragonal $\it{c}$-axis between Cu-O
bi-layers may be partially blocked by the magnetic dipole field
B$_{dipole}$ of ordered Ru moments in the Ru-O layer.\cite{p47}

\begin{figure}
\includegraphics{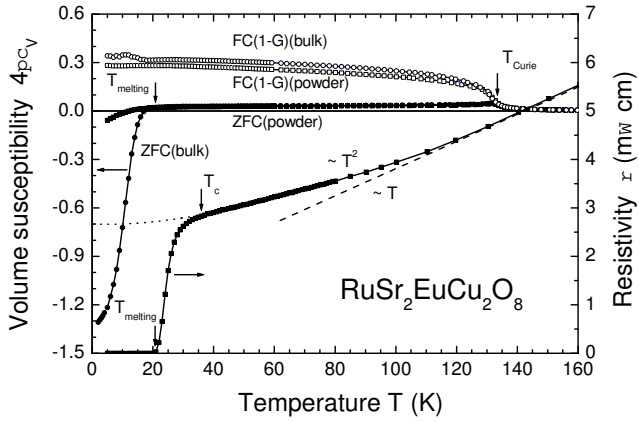}
\caption{\label{fig2}  The electrical resistivity $\rho$(T) and
volume magnetic susceptibility 4$\pi\chi_{V}$(T) in 1-G field-cooled
(FC) and zero-field-cooled (ZFC) modes for oxygen-annealed bulk and
powder RuSr$_{2}$EuCu$_{2}$O$_{8}$ samples.}
\end{figure}

The Meissner shielding at 2 K is complete (4$\pi\chi_{V}$ =
4$\pi$M/B$_{a} \sim$ 1.3) for ZFC bulk sample, but much reduced
(-0.1) in the powder sample. However, in 1-G FC mode, no such an
effect can be detected below T$_{melting}$ due to strong flux
pinning.

Low-field ($\pm$20 G) superconducting hysteresis loop at 2 K for
bulk sample RuSr$_{2}$EuCu$_{2}$O$_{8}$ and
RuSr$_{2}$GdCu$_{2}$O$_{8}$ as reference are shown in Fig. 3. The
initial magnetization curve deviates from straight line at 2 G and 3
G for the Eu- and Gd-compound, respectively. The narrow region of
full Meissner effect roughly reflects the temperature-dependent
lower critical field in the $\it{ab}$-plane B$_{c1}^{ab}$(T). The
average lower critical field B$_{c1}^{ave}$ for bulk sample as
determined from the peak of initial diamagnetic magnetization curves
is 7 G for R = Eu and 13 G for R = Gd. The effect on the exact peak
value due to the surface barrier pinning is neglected. For
RuSr$_{2}$EuCu$_{2}$O$_{8}$, B$_{c1}^{ave}$ decreases steadily from
7 G at 2 K to 6 G at 5 K, 4 G at 10 K, and below 1 G at 15 K. A
simple empirical parabolic fitting gives B$_{c1}^{ave}$(T) =
B$_{c1}^{ave}$(0)[1 - (T/T$_{SVS}$)$^{2}$], with average
B$_{c1}^{ave}$(0) $\sim$ 7 G and spontaneous vortex state
temperature T$_{SVS}$ = 16 K. The Ginzburg-Landau anisotropy formula
B$_{c1}^{ave}$ = (2B$_{c1}^{ab}$ + B$_{c1}^{c}$)/3, then provides an
estimated $\it{c}$-axis lower critical field B$_{c1}^{c}$) $\sim$ 17
G and anisotropy parameter $\sim$ 8.5.

\begin{figure}
\includegraphics{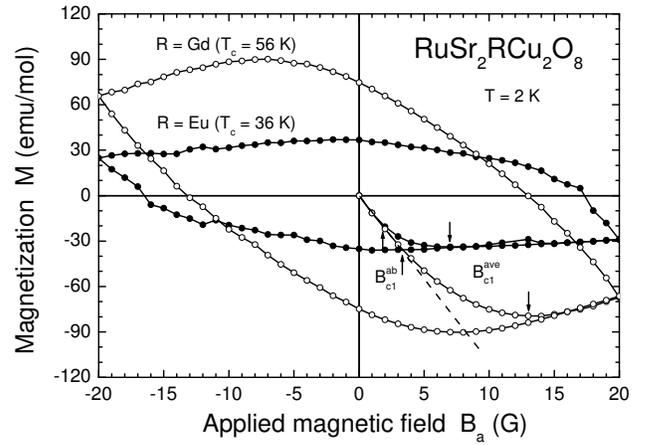}
\caption{\label{fig3}  The low-field superconducting hysteresis
loops M-B$_{a}$ at 2 K for RuSr$_{2}$GdCu$_{2}$O$_{8}$ and
RuSr$_{2}$EuCu$_{2}$O$_{8}$. Average lower critical field
B$_{c1}$(ave) at peak values and $\it{ab}$-plane B$_{c1}^{ab}$ for
deviation from initial linear lines are indicated by arrows.}
\end{figure}

\begin{figure}
\includegraphics{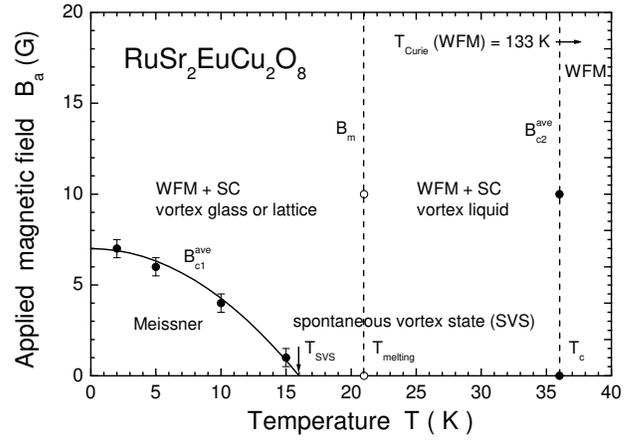}
\caption{\label{fig4} The low field, low temperature superconducting
phase diagram B$_{a}$(T) of RuSr$_{2}$EuCu$_{2}$O$_{8}$. The
spontaneous vortex state (SVS) occurrs between T$_{SVS}$ = 16 K and
T$_{c}$ = 36 K. Vortex lattice/glass melting temperature
T$_{melting}$ is defined from temperature at which resistivity drops
to zero.}
\end{figure}

The lower field superconducting phase diagram for the
polycrystalline bulk sample is shown in Fig. 4. The average lower
critical field B$_{c1}^{ave}$ separates the Meissner state and
vortex state. The upper critical field B$_{c2}$ and vortex melting
field B$_{melting}$ determinated from magnetoresistivity
measurements are field-independent below 20 G. The WFM-induced
internal dipole field B$_{dipole}$ of 8.8 G on the CuO$_{2}$
bi-layers is estimated using extrapolated B$_{c1}^{ave}$ value at T
= 0, (B$_{c1}^{ave}$(0) + B$_{dipole}$)/B$_{c1}^{ave}$(0) =
T$_{c}$/T$_{SVS}$. It further yields a small net spontaneous
magnetic moment $\mu_{s}$ of 0.1 $\mu_{B}$ per Ru, based on the
relation of B$_{dipole} \sim$ 2$\mu_{s}$/($\it{c}$/2)$^{3}$, where
$\it{c}$/2 = 0.58 nm is the distance between midpoint of CuO$_{2}$
bi-layers and two nearest-neighbor Ru moments. If the WFM structure
is indeed a G-type antiferromagnetic order with 1.5 $\mu_{B}$ for
Ru$^{5+}$ in t$_{2g}$ states canted along the tetragonal basal
plane, the small $\mu_{s}$ would give a canting angle of 4$^{o}$
from the tetragonal $\it{c}$-axis and be difficult to be detected in
neutron diffraction with a resolution $\sim$ 0.1 $\mu_{B}$.

\begin{figure}
\includegraphics{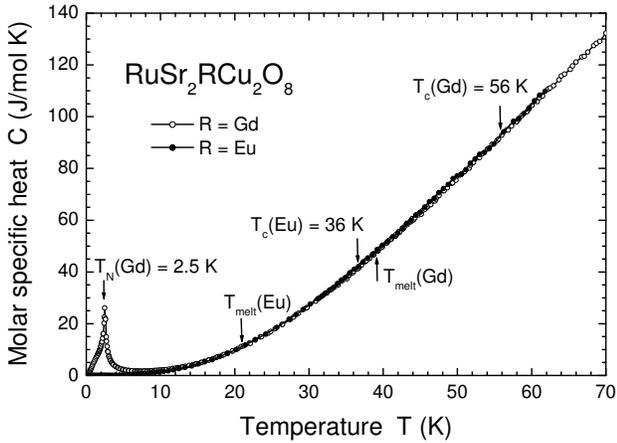}
\caption{\label{fig5}  The molar specific heat of
RuSr$_{2}$RCu$_{2}$O$_{8}$ (R = Eu, Gd). Antiferromagnetic Gd$^{3+}$
ordering prevails at 2.5 K.}
\end{figure}

The molar specific heat data up to 70 K in Fig. 5 show a good
agreement between Eu- and Gd-compounds, except that a peak reflects
the antiferromagnetic Gd$^{3+}$ ordering near T$_{N} \sim$ 2.5 K.
Consistent with previous results for lower-T$_{c}$ Gd-compounds in
zero applied magnetic field.\cite{p28,p15} No visible jump in
specific heat was observed near T$_{c}$ = 36 K. This is not
surprising, since only the electronic component in specific heat
would change with superconducting transition, but it is easily
overshadowed by the much larger phonon contribution. Specifically,
assuming a same magnitude as that observed in
La$_{1.85}$Sr$_{0.15}$CuO$_{4}$ ($\Delta$C $\sim$ 0.33 J/mol K at
T$_{c}$ = 37 K) and YBa$_{2}$Cu$_{3}$O$_{7}$ ($\Delta$C $\sim$ 4.6
J/mol K at T$_{c}$ = 92 K),\cite{p53} an estimated $\Delta$C $\sim$
1 J/mol K at T$_{c}$ here is only about 1$\%$ of total specific
heat, falling below the experimental precision. In addition, the
broad resistivity transition due to vortex melting would further
points to a correspondingly reduced height of $\Delta$C.

It would be of interest to obtain information on the Gd$^{3+}$
ordering. To do so, delineation of various contributions to the
total specific heat begins with the nonmagnetic Eu-compound up to 7
K. In the format of C/T versus T$^{2}$, the data in Fig. 6 can be
well fitted by the sum of four terms with different temperature
dependence:
\begin{equation}
C = \beta T^{3} + \alpha T^{2} + \gamma T + \frac{\eta}{T^{2}}.
\end{equation}
\noindent

The coefficient of the first term, $\beta$ = 0.89 mJ/mol K$^{4}$,
can be used to derive a Debye temperature $\theta_{D}$ of the
lattice,
\begin{equation}
\beta = n(12\pi^{4}/5)N_{A}\it{k}/\theta_{D}^{3},
\end{equation}
\noindent where $\it{N_{A}}$ is Avogadro's number, $\it{k}$ the
Boltzmann constant, and the number of atoms per formula unit n = 14.
The $\theta_{D}$ value of 312 K thus obtained supports the validity
of the T$^{3}$-dependence approximation in Debye model for the
lattice specific heat below 7 K $\sim \theta_{D}$/50. The quadratic
term has two possible sources: the nodal line excitation for
$\it{d}$-wave pairing symmetry and the spin wave excitation of WFM
Ru sublattice. The fact that the observed $\alpha$ value of 4.2
mJ/mol K is much large than 0.1 mJ/mol K of YBa$_{2}$Cu$_{3}$O$_{7}$
could be an indication of a less important nodal line excitation,
but an enhanced spin wave excitation. The linear term is considered
normally as an electronic contribution, which is not expected to
exist in a superconductor at temperature much lower than T$_{c}$.
While the observed coefficient $\gamma$ = 7.3 mJ/mol K$^{2}$ is
comparable to that of some cuprates, its origin remains to be
identified. One plausible explanation is based on the complicated
magnetic structure and mixed valence. Such a scenario could lead to
a spin glass-like lattice, for which an even larger linear term in
specific heat has been observed in another Ru compound of
Ba$_{2}$PrRuO$_{6}$. \cite{p54}

\begin{figure}
\includegraphics{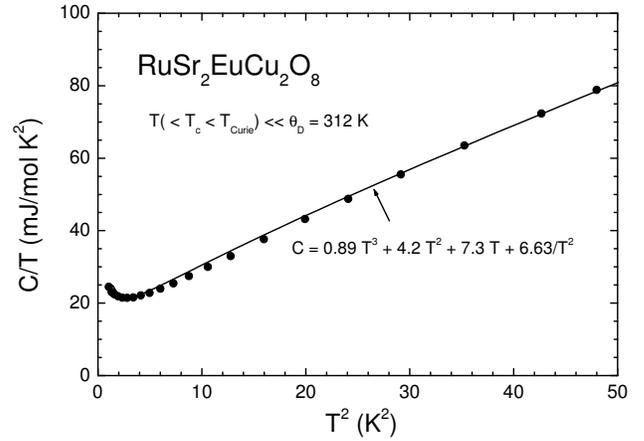}
\caption{\label{fig6}  Low temperature C/T versus T$^{2}$ of
RuSr$_{2}$EuCu$_{2}$O$_{8}$ from 1 K to 7 K. Data above 1 K can be
fitted using C(T) = $\beta$T$^{3}$ + $\alpha$T$^{2}$ + $\gamma$T +
$\eta$/T$^{2}$ with Debye temperature $\theta_{D}$ = 312 K.}
\end{figure}

The last term with a T$^{-2}$ dependence is most likely the
high-temperature tail of a Schottky anomaly. Its occurrence at the
relatively low temperatures suggests nuclear energy splittings being
the cause. Such energy splittings occur typically for nuclei having
a spin I and magnetic moment $\mu_{n}$ in a hyperfine magnetic field
H$_{hf}$. For the calorimetrical measurements under consideration,
they are is most likely associated with the Ru nuclei, since the
4$\it{d}$ magnetic moments of ordered Ru ions are spatially fixed,
polarizing the $\it{s}$-electrons and producing a net spin at the
nuclei, yielding a hyperfine field. There are two Ru isotopes with
non-zero $\mu_{n}$: $^{99}$Ru (fractional natural abundance A =
0.1276, I = 5/2, and $\mu_{n}$ = -0.6413) and $^{101}$Ru (A =
0.1706, I = 5/2, and $\mu_{n}$ = -0.7188).\cite{p55} However,
nuclear energy splittings can also be caused by the interaction
between the qudrupole moment of a nucleus and the electric field
gradient produced by neighboring atoms. The electric field gradient
could be quite high in the layered compound. Meanwhile, Cu  and Eu
or $^{155}$Gd (A = 14.7$\%$) and $^{157}$Gd (A = 15.7$\%$) nuclei
all have non-zero quadrupole moment. Without the full knowledge of
magnetic hyperfine field and electric field gradient, it is not
feasible at present to delineate the experimentally obtained $\eta$
of 6.63 mJ K/mol into the two different contributions.

\begin{figure}
\includegraphics{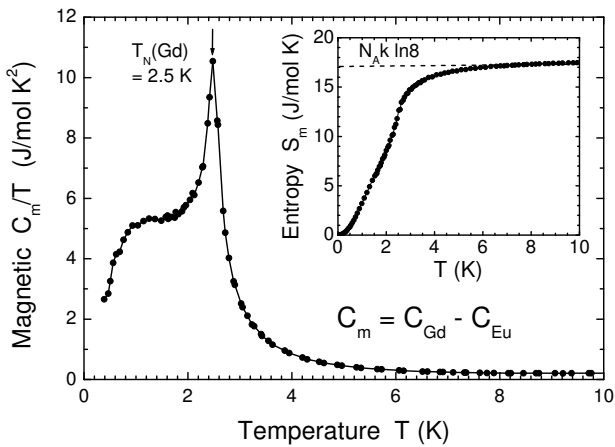}
\caption{\label{fig7}  Temperature dependence of magnetic specific
heat and entropy (inset) associated with Gd$^{3+}$ ordering in
RuSr$_{2}$GdCu$_{2}$O$_{8}$.}
\end{figure}

By assuming that its various coefficients in Eq. (1) for Eu-compound
remain the same for the Gd-compound, One can then obtain the
magnetic contribution to specific heat associated with
antiferromagnetic Gd$^{3+}$ ordering as
\begin{equation}
C_{m} = C_{Gd} - C_{Eu}.
\end{equation}
\noindent The results are shown in Fig. 7. Using the format of
C$_{m}$/T versus T. It is of interest to note a broad shoulder below
T$_{N}$, a common feature seemingly prevailing in other similar type
of compounds such as GdBa$_{2}$Cu$_{3}$O$_{7}$,
GdBa$_{2}$Cu$_{4}$O$_{8}$ and TlBa$_{2}$GdCu$_{2}$O$_{7}$.
\cite{p56,p57,p58} According to Fishman and Liu,\cite{p59} it is due
to spin fluctuations in the normally ordered state, and such
fluctuations are more pronounced for large spins. Indeed, Gd$^{3+}$
has the largest spin among all R$^{3+}$ ions. The areal integral in
Fig. 7, including that associated with the broad shoulder should
yield the magnetic entropy,
\begin{equation}
S_{m} = \int(C_{m}/T)dT.
\end{equation}
\noindent  As shown in the inset, S$_{m}$ reaches a saturation value
of 17.6 J/mol K around 10 K. Considering the built-in approximation
in Eq. (4), it agrees exceptional well with the theoretical value of
$\it{N_{A}k}$ ln(2J+1) = $\it{N_{A}k}$ ln8 = 17.2 J/mol K for the
complete ordering of Gd$^{3+}$.
\newline
\newline

\section{conclusion}
The lower critical field with B$_{c1}$(0) = 7 G and T$_{SVS}$ = 16 K
indicates the existence of a spontaneous vortex state (SVS) between
16 K and T$_{c}$ of 36 K. This SVS state is closely related to the
weak-ferromagnetic order with a net spontaneous magnetic moment of
$\sim$ 0.1 $\mu_{B}$/Ru, which generates a weak magnetic dipole
field around 8.8 G in the CuO$_{2}$ bi-layers. The vortex melting
transition temperature at 21 K obtained from resistivity
measurements and the onset of diamagnetic signal indicates a broad
vortex liquid region due to the coexistence and interplay between
superconductivity and WFM order. No visible specific heat jump was
observed near T$_{c}$ for Eu- and Gd-compound, since the electronic
specific heat is easily overshadowed by the large phonon
contributions and the expected jump would spread over a wide range
of temperature due to vortex melting. Finally, the magnetic entropy
associated with Gd$^{3+}$ antiferromagnetic ordering at 2.5 K is
confirmed to be close to $\it{N_{A}k}$ ln8 for J = S = 7/2.

\begin{acknowledgments}
This work was supported by the National Science Council of R.O.C.
under contract Nos. NSC95-2112-M-007-056-MY3 and
NSC95-2112-M-032-002.
\end{acknowledgments}



\begin{thebibliography}{99}

\bibitem{p1} L. Bauernfeind, W. Widder, and H. F. Braun,
Physica C {\bf 254}, 151 (1995).

\bibitem{p2} L. Bauernfeind, W. Widder, H. F. Braun,
J. Low Temp. Phys. {\bf 105}, 1605 (1996).

\bibitem{p3} K. B. Tang, Y. T. Qian, L. Yang, Y. D. Zhao, Y. H. Zhang,
Physica C {\bf 282-287}, 947 (1997).

\bibitem{p4} J. L. Tallon, C. Bernhard, M. Bowden, P. Gilberd, T. Stoto, and D. Pringle,
IEEE Trans. Appl. Supercon. {\bf 9}, 1696 (1999).

\bibitem{p5} C. Bernhard, J. L. Tallon, Ch. Niedermayer, Th. Blasius, A. Golnik, E. Brucher, R. K. Kremer, D. R. Noakes, C. E. Stronach, and E. J. Ansaldo,
Phys. Rev. B {\bf 59}, 14099 (1999).

\bibitem{p6} A. C. McLaughlin, W. Zhou, J. P. Attfield, A. N. Fitch, and J. L. Tallon,
Phys. Rev. B {\bf 60}, 7512 (1999).

\bibitem{p7} J. L. Tallon, J. W. Loram, G. V. M. Williams, and C. Bernhard,
Phys. Rev. B {\bf 61}, R6471 (2000).

\bibitem{p8} C. Bernhard, J. L. Tallon, E. Brucher, and R. K. Kremer,
Phys. Rev. B {\bf 61}, R14960 (2000).

\bibitem{p9} J. W. Lynn, B. Keimer, C. Ulrich, C. Bernhard, and J. L. Tallon,
Phys. Rev. B {\bf 61}, R14964 (2000).

\bibitem{p10} O. Chmaissem, J. D. Jorgensen, H. Shaked, P. Dollar, and J. L. Tallon,
Phys. Rev. B {\bf 61}, 6401 (2000).

\bibitem{p11} G. V. M. Williams, and S. Kramer,
Phys. Rev. B {\bf 62}, 4132 (2000).

\bibitem{p12} C. W. Chu, Y. Y. Xue, S. Tsui, J. Cmaidalka, A. K. Heilman, B. Lorenz, and R. L. Meng,
Physica C {\bf 335}, 231 (2000).

\bibitem{p13} A. C. Mclaughlin, V. Janowitz, J. A. McAllister, and J. P. Attfield,
Chem. Commun. {\bf 2000}, 1331 (2000).

\bibitem{p14} X. H. Chen, Z. Sun, K. Q. Wang, Y. M. Xiong, H. S. Yang, H. H. Wen, Y. M. Ni, Z. X. Zhao,
J. Phys. Cond. Mat. {\bf 12}, 10561 (2000).

\bibitem{p15} J. L. Tallon, J. W. Loram, G. V. M. Williams, and C. Bernhard, Phys. Rev. B
{\bf 61}, R6471 (2000).

\bibitem{p16} R. L. Meng, B. Lorenz, Y. S. Wang, J. Cmaidalka, Y. Y. Xue, and C. W. Chu,
Physica C {\bf 353}, 195 (2001).

\bibitem{p17} V. P. S. Awana, J. Nakamura, M. Karppinen, H. Yamauchi, S. K. Malik, and W. B. Yelon,
Physica C {\bf 357-360}, 121 (2001).

\bibitem{p18} D. P. Hai, S. Kamisawa, I. Kakeya, M. Furuyama, T. Mochiku, and K. Kadowaki,
Physica C {\bf 357-360}, 406 (2001).

\bibitem{p19} A. P. Litvinchuk, S. Y. Chen, M. N. Ilive, C. L. Chen, C. W. Chu, and V. N. Popov,
Physica C {\bf 361}, 234 (2001).

\bibitem{p20} C. T. Lin, B. Liang, C. Ulrich, C. Berhard,
Physica C {\bf 364-365}, 373 (2001).

\bibitem{p21} H. Takagiwa, J. Akimitsu, H. Kawano-Furukawa, and H. Yoshizawa,
J. Phys. Soc. Jpn. {\bf 70}, 333 (2001).

\bibitem{p22} J. D. Jorgensen, O. Chmaissem, H. Shaked, S. Short, P. W. Klamut, B. Dabrowski, and J. L. Tallon,
Phys. Rev. B {\bf 63}, 054440 (2001).

\bibitem{p23} R. S. Liu, L.-Y. Jang, H.-H Hung, and J. L. Tallon,
Phys. Rev. B {\bf 63}, 212507 (2001).

\bibitem{p24} M. Pozek, A. Dulcic, P. Paar, G. V. M. Williams, and S. Kramer,
Phys. Rev. B {\bf 64}, 064508 (2001).

\bibitem{p25} V. G. Hadjiew, J. Backstrom, V. N. Popov, M. N. Iliev, R. L. Meng, Y. Y. Xue, and C. W. Chu,
Phys. Rev. B {\bf 64}, 134304 (2001).

\bibitem{p26} Y. Tokunaga and H. Kotegawa and K. Ishida and Y. Kitaka and H. Takagiwa and J. Akimitsu,
Phys. Rev. Lett. {\bf 86}, 5767 (2001).

\bibitem{p27} P. W. Klamu, B. Dabrowski, S. Kolesnik, M. Maxwell, and J. Mais,
Phys. Rev. B {\bf 63}, 224512 (2001).

\bibitem{p28} X. H. Chen, Z. Sun, K. Q. Wang, S. Y. Li, Y. M. Xiong,
M. Yu, and L. Z. Cao, Phys. Rev. B {\bf 63}, 64506 (2001).

\bibitem{p29} B. Lorenz, Y. Y. Xue, R. L. Meng, and C. W. Chu,
Phys. Rev. B {\bf 65}, 174503 (2002).

\bibitem{p30} T. P. Papageorgiou, H. F. Braun, and T. Herrmannsdorfer,
Phys. Rev. B {\bf 66}, 104509 (2002).

\bibitem{p31} H. Fujishiro, M. Ikebe, and T. Takahashi,
J. Low Temp. Phys. {\bf 131}, 589 (2003).

\bibitem{p32} C. Shaou, H. F. Braun, and T. P. Papageorgiou,
J. Alloys and Compounds {\bf 351}, 7 (2003).

\bibitem{p33} A. Vecchione, M. Gombos, C. Tedesco, A. Immirzi, L. Marchese, A. Frache, C. Noce, and S. Pace,
Intern. J. Mod. Phys. B {\bf 17}, 899 (2003).

\bibitem{p34} F. Cordero, M. Ferretti, M. R. Cimberle, and R. Masini,
Phys. Rev. B {\bf 67}, 144519 (2003).

\bibitem{p35} H. Sakai, N. Osawa, K. Yoshimura, M. Fang, and K. Kosuge,
Phys. Rev. B {\bf 67}, 184409 (2003).

\bibitem{p36} Y. Y. Xue, F. Chen, J. Cmaidalka, R. L. Meng, and C. W. Chu,
Phys. Rev. B {\bf 67}, 224511 (2003).

\bibitem{p37} J. E. McCrone, J. L. Tallon, J. R. Cooper, A. C. MacLaughlin, J. P. Attfield, and C. Bernhard,
Phys. Rev. B {\bf 68}, 064514 (2003).

\bibitem{p38} A. Lopez, I. S. Azevedo, J. E. Musa, and E. Baggio-Saitovitch,
Phys. Rev. B {\bf 68}, 134516 (2003).

\bibitem{p39} S. Garcia, J. E. Musa, R. S. Freitas, and L. Ghivelder,
Phys. Rev. B {\bf 68}, 144512 (2003).

\bibitem{p40} T. P. Papageorgiou, H. F. Braun, T. Gorlach, M. Uhlarz, and H. v. Lohneysen,
Phys. Rev. B {\bf 68}, 144518 (2003).

\bibitem{p41} V. P. S. Awana, T. Kawashima, and E. Takayama-Muromachi,
Phys. Rev. B {\bf 67}, 172502 (2003).

\bibitem{p42} P. W. Klamut, B. Dabrowski, S. M. Mini, M . Maxwell, J. Mais, I. Felner, U. Asaf, F. Ritter, A. Shengelaya, R. Khasanov, I . M. Savic, H. Keller, A. Wisniewski, R. Puzniak, I. M. Fita, C. Sulkowski, and M. Matusiak,
Physica C {\bf 387}, 33 (2003).

\bibitem{p43} T. Nachtrab, D. Koelle, R. Kleiner, C. Bernhard, and C. T. Lin,
Phys. Rev. Lett. {\bf 92}, 117001 (2004).

\bibitem{p44} Y. Y. Xue, B. Lorenz, A. Baikalov, J. Cmaidaka, F. Chen, R. L. Meng, and C. W. Chu,
Physica C {\bf 408-410}, 638 (2004).

\bibitem{p45} S. Garcia and L. Ghivelder,
Phys. Rev. B {\bf 70}, 052503 (2004).

\bibitem{p46} C. J. Liu., C. S. Sheu, T. W. Wu, L. C. Huang, F. H. Hsu, H. D. Yang, G. V. M. Williams, and C. C. Liu,
Phys. Rev. B {\bf 71}, 014502 (2004).

\bibitem{p47} C. Y. Yang, B. C. Chang, H. C. Ku, and Y. Y. Hsu,
Phys. Rev. B {\bf 72}, 174508 (2005).

\bibitem{p48} T. P. Papageorgiou, E. Casini, H. F. Braun, T. Herrmannsdorfer, A. D. Bianchi, and J.
Wosnitza, Euro. Phys. J. B {\bf 52}, 383 (2006).

\bibitem{p49} Y. C. Lin, T. Y. Chiu, M. F. Tai, B. N. Lin, P. C. Guan, and H. C. Ku,
Intern. J. Mod. Phys. B {\bf 19}, 339 (2005).

\bibitem{p50} T. Y. Chiu, Y. C. Lin, M. F. Tai, B. N. Lin, P. C. Guan, B. C. Chang, and H. C. Ku,
Chin. J. Phys. {\bf 43}, 616 (2005).

\bibitem{p51} H. C. Ku, C. Y. Yang, B. C. Chang, B. N. Lin, Y. Y. Hsu, and M. F. Tai,
J. Appl. Phys. {\bf 97}, 10B110 (2005).

\bibitem{p52} B. C. Chang, C. Y. Yang, Y. Y. Hsu, B. N. Lin, and H. C. Ku
AIP conference proceeding {\bf 850}, 677 (2006).

\bibitem{p53} Charles P. Jr., Horacio A. Farach, Richard J. Creswick {\it Superconductivity}
(Academic Press, Inc. San Diego, 1995), chap. 4.

\bibitem{p54} S. M. Rao, M. K. Wu, J. K. Srivastava, B. H. Mok, C. Y. Lu, Y. C. Liao,
Y. Y. Hsu, Y. S. Hsiue, Y. Y. Chen, S. Neeleshwar, S. Tsai, J. C.
Ho, and H. L. Liu, Phys. Lett. A {\bf 324}, 71 (2004).

\bibitem{p55} CRC Handbook of Chemistry and Physics, 86th ed., Taylor and Francis,
2005-2006.

\bibitem{p56} S. E. Brown, J. D. Thompson, J. O. Willis, R. M. Aikin, E. Zirngiebl,
J. L. Smith, Z. Fisk, and R. B. Schwarz, Phys. Rev. B. {\bf 36},
2298 (1987).

\bibitem{p57} J. C. Ho, Y. Y. Chen, Y. D. Yao, W. S. Huang, S. R. Sheen, J. C. Huang,
and M. K. Wu, Physica C {\bf 282-287}, 1403 (1997).

\bibitem{p58} Y. Y. Chen, C. C. Lai, B. S. Chiou, J. C. Ho, and H. C. Ku, Phys. Rev. B
{\bf47}, 12178 (1993).

\bibitem{p59} R. S. Fishman and S. H. Liu, Phys. Rev. B {\bf 40}, 11028 (1989).


\end{thebibliography}
\end{document}